# The First-principles Study on the Mechanics, Optical and Phonon Properties of Carbon Chains

Jin-Ping Li（李金平）[1 ¤*], Song-He Meng（孟松鹤）[1], Han-Tao Lu（陆汉涛）[2]
and Takami Tohyama（遠山貴巳）[3]

(1 Center for Composite Materials and Structure, Harbin Institute of Technology, Harbin 150080, China)

(2 Center for Interdisciplinary Studies & Key Laboratory for Magnetism and Magnetic Materials of the MoE, Lanzhou University, Lanzhou 730000, China)

(3 Department of Applied Physics, Tokyo University of Science, Tokyo 125-8585, Japan)

**Abstract:** Besides graphite, diamond, graphene, carbon nanotubes, and fullerenes etc., there is another allotrope of carbon: carbyne, existing in the form of a one-dimensional chain of carbon atoms. It has been theoretically predicted that carbyne would be stronger, stiffer and more exotic than other materials that have been synthesized before. In this article, the two kinds of carbyne, i.e., cumulene and polyyne, are investigated by the first principles, where the mechanical properties, electronic structure, optical and phonon properties of the two carbynes are calculated. The results on the crystal binding energy and the formation energy show that polyyne is more stable and harder than cumulene, and both are difficult to be synthesized from diamond or graphite.  The tensile stiffness, bond stiffness and Young's modulus of cumulene are 94.669 eV/Å, 90.334GPa and 60.62GPa, respectively; while the corresponding values of polyyne are 94.939eV/Å, 101.42GPa and 60.06GPa, respectively. The supercell calculation shows that carbyne is the most stable at N=5, where N is the supercell number. It indicates that the carbon chain with 10 atoms is the most stable. The calculation on the electronic band structure shows that cumulene is conductor, while polyyne is semiconductor with a band gap as 0.37eV.  The dielectric function of carbynes varies in different direction, consistent with the one dimensional nature of the carbon chains. In the phonon dispersion of cumulene there are imaginary frequencies with the lowest value down to -3.817THz, which indicates that cumulene could be unstable at room temperature and normal pressure.

---

¤Project supported by the National Natural Science Foundation of China (Grant Nos. 11672087, 11474136).

* Corresponding author: lijinping@hit.edu.cn

# 1 Introduction

A new type of carbon chains, called carbyne [1], has been received much attention in the community of material because of its unique mechanical and electronic properties and potential value for applications, e.g., in nanoelectronic or spintronic devices [2-7] and hydrogen storage [8]. In general, carbyne is a chain of carbon atoms that are linked either by alternate triple and single bonds (-C≡C-), i.e., polyyne, or by consecutive double bonds (=C=C=), i.e., cumulene [9]. Carbyne is something of a mystery, since astronomers believe that they have detected its signature in interstellar space; but chemists, on the other hand, have been bickering for decades over whether they had ever created this stuff on Earth [10]. A couple of years ago, however, chemists have synthesized carbyne chains successfully up to 44 atoms long in solution [11], which comprises an important step towards the realistic application of the material.

The possibility of the realization of stable carbynes in labs stimulates considerable theoretical interest and investigations. Some interesting features of carbynes have been revealed and discussed recently from the theoretical side, e.g., it could be stronger, stiffer and more exotic than other known materials [12]. However, to the best of our knowledge, the need of a fully systematic investigation of the physical properties of carbynes is still to be met. In this paper, by means of the first-principles calculations we systematically study the mechanical, electronic structure, optical and phonon properties of the two carbynes. The comparison with other numerical results is also provided. This work will help us to understand caybyne in order to make good use of it in the future.

# 2. Computational methodology

First-principles calculations are performed with plane-wave ultrasoft pseudopotential by means of GGA with PW91 functional as implemented in the CASTEP code (Cambridge Sequential Total Energy Package) [13]. The ionic cores are represented by ultrasoft pseudopotential for C atoms. For C atom, the configuration is [He]$2s^22p^2$, where the $2s^2$and $2p^2$ electrons are explicitly treated as valence electrons. The plane-wave cut-off energy is 380eV, and the Brillouin-zone integration is performed over the $3\times3\times10$ grid sizes using the Monkorst-Pack method for structure optimization. This set of parameters assures the total energy convergence of $5.0\times10^{-6}$eV/atom, the maximum force of 0.01eV/Å, the maximum stress of 0.02 GPa and the maximum displacement of $5.0\times10^{-4}$Å.

In the following sections, we perform calculations on carbynes after or without having optimized the geometry structure by the GGA-PW91.

## 3. Crystal binding energy and formation energy

Crystal binding energy $W$ is defined as the energy difference (per atom) between the total energy $E_0$ when the crystalline is stable for $N$ atoms, and the total energy of the $N$ atoms $E_N$ when they are in a free state, i.e., $W= (E_N-E_0)/N$. For covalent crystals, when the crystal binding energy is larger, the crystal will be more stable, usually with higher melting-point and hardness. The energy of a free carbon atom is calculated to be -147.017951eV. For cumulene, $E_N$ is given by the total energy of 2 free carbon atoms, while $E_0$ for 2 carbon atoms is estimated to be -308.375891eV. Thus, the crystal binding energy reads $W$=[(-147.017951)×2-(-308.375891)]/2=7.1699945eV. On the other hand, for polyyne $E_0$=-308.378995eV, leading to $W$=7.1715465eV. The values of $W$ are consistent with previously reported values of $W$=6.99~8.19eV [7]. From the crystal binding energy, we conclude that both cumulene and polyyne can be stable against free carbons. Comparatively, polyyne is more stable than cumulene with the gain of 1.55 meV per atom. This is consistent with the fact that the unit structure of cumulene (=C=C=) is known to undergo a Peierls transition into that of polyyne (-C≡C-) [9].

The stability of carbynes can be further examined by comparing the formation energy $E_f$ with those of other carbon allotropes, i.e., diamond and graphite. The total energy of diamond and graphite for two carbon atoms are calculated to be -309.873429eV and -310.129136eV, respectively. We can see that they are lower than $E_f$ for both cumulene and polyyne, which underlines the difficulty of synthesizing carbynes from diamond or graphite. The significant energy difference also indicates a possible instability of carbynes at room temperatures.

## 4. Mechanical properties

## 4.1 Cumulene

### 4.1.1 Tensile stiffness

The most basic mechanical property of the carbyne chain is its tensile stiffness, which is defined as

$$C = \frac{1}{a}\frac{\partial^2 E}{\partial \varepsilon^2} \tag{1}$$

where $a$ is the unit cell length (2.565 Å), $E$ is the strain energy per two C atoms, and ε is the strain. As shown in Fig. 1, fitting our theoretical data to equation (1) by a second-order polynomial (Fig.1) yields a tensile stiffness of $C$ = 94.669 eV/Å, which is in agreement with earlier work, $C$ = 95.56 eV/Å [12].

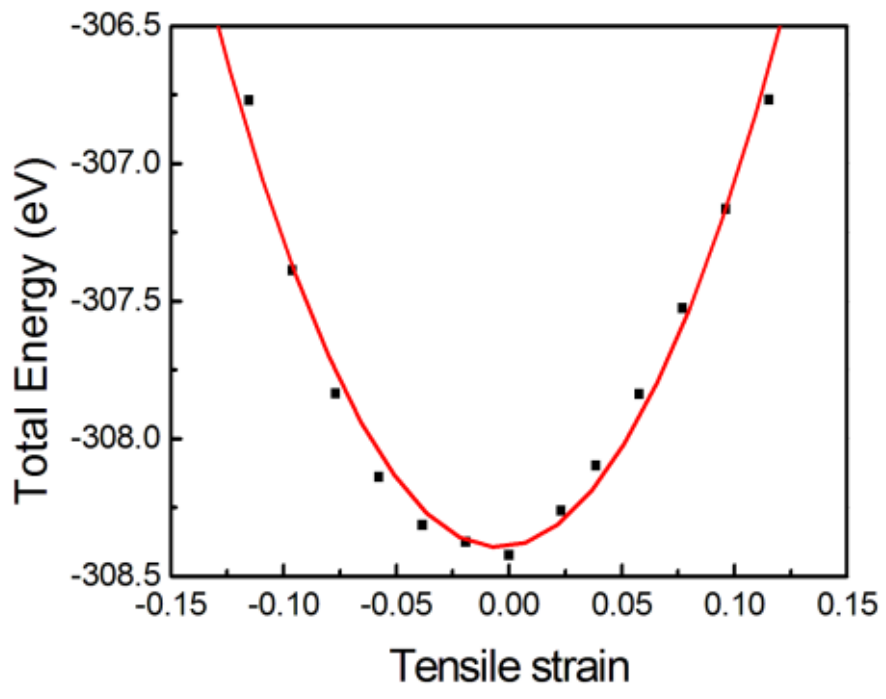

Fig.1. The dependence of the total energy on stretch strain for cumulene. Black dots are numerical results, and the red curve is a fitting curve with second-order polynomial.

### 4.1.2 Bond stiffness

Another mechanical property is the bond stiffness, which describes the change of the bond length between neighboring $C$ atoms ($d$) with respect to an applied pressure ($P$). Taking the bond length at ambient pressure $P_0$ to be $d_0$, we expand the relative bond length ($d/d_0$) up to second order of $P$,

$$d/d_0 = C_0 + C_1 P + C_2 P^2 \tag{2}$$

and then determine the coefficients by fitting the calculated $d/d_0$ to the polynomial. The chemical bond stiffness $K$ is given by

$$K = \left|\frac{d(d/d_0)}{dP}\right|^{-1} = |C_1 + 2C_2 P|^{-1} \tag{3}$$

Figures 2(a) and 2(b) show the first-principles results of the relative bond length $d/d_0$ (in dots) and the bulk modulus $B$ (in dots) as a function of hydrostatic pressure, respectively, which reflects the bond stiffness of cumulene. The red curves are produced by a second-order fitting. We obtain the chemical bond stiffness $K$ in cumulene to be 90.334GPa in Fig. 2(a). In Fig. 2(b), we can observe that the bulk modulus $B$ increases with the pressure. When $P$ is about 1GPa, $B$ reaches 1531.4GPa. The above data shows that cumulene is very stiff even among the carbon materials, e.g., the bulk modulus of diamond is about 600GPa, and the value of cumulene, on the other hand, more than 1500GPa.

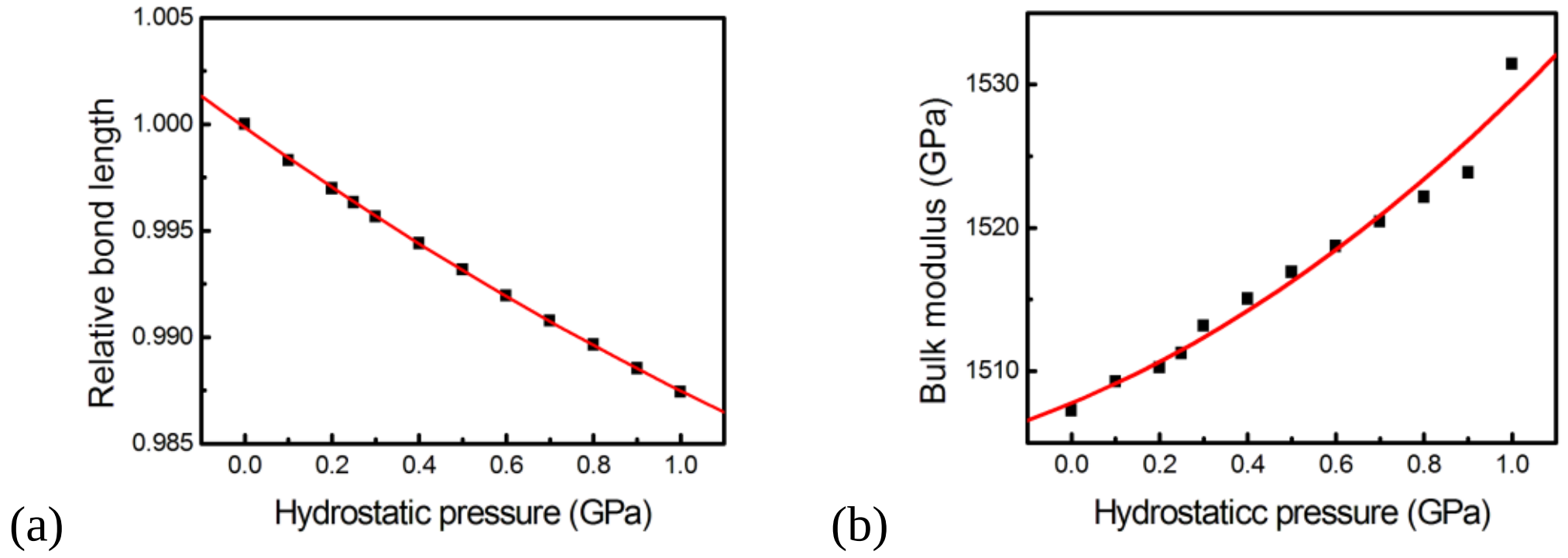

Fig.2 The relative bond length $d/d_0$ (a) (in (a)) and the bulk modules $B$ (b) (in (b)) as a function of hydrostatic pressure for cumulene.

### 4.1.3 The relation between stress and strain

The relation between stress and strain in cumulene can be studied from two aspects, i.e., either by studying the response of the bond length to different loads applied to the chain (as in Fig.3), or by determining the consequent stress values following the change of the bond length (as in Fig.4). It can be observed in Fig.3 that the strain increases almost linearly with stress up to 2.5GPa, while the linearity of the decrease of the bulk modulus with the tensile stress is also plain in Fig.3(b). From Fig.4, it is easily to see the relationship between strain and stress, that is to say, the tensile stress increases with bond length of cumulene.

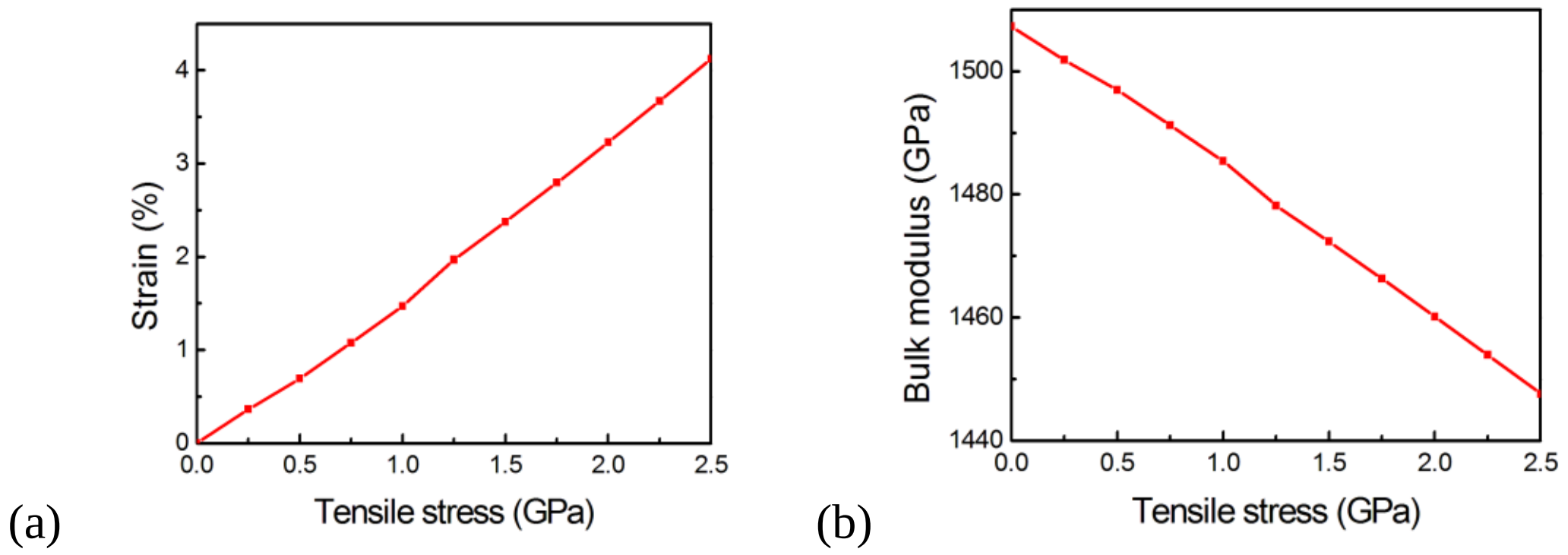

Fig.3 The strain (a) and bulk modulus (b) of cumulene as a function of tensile stress.

The Young's modulus of cumulene can be calculated from Fig.3(a) as

$$E = \frac{\sigma}{\varepsilon} = \frac{2.5GPa}{4.124\%} = 60.62GPa \quad (4)$$

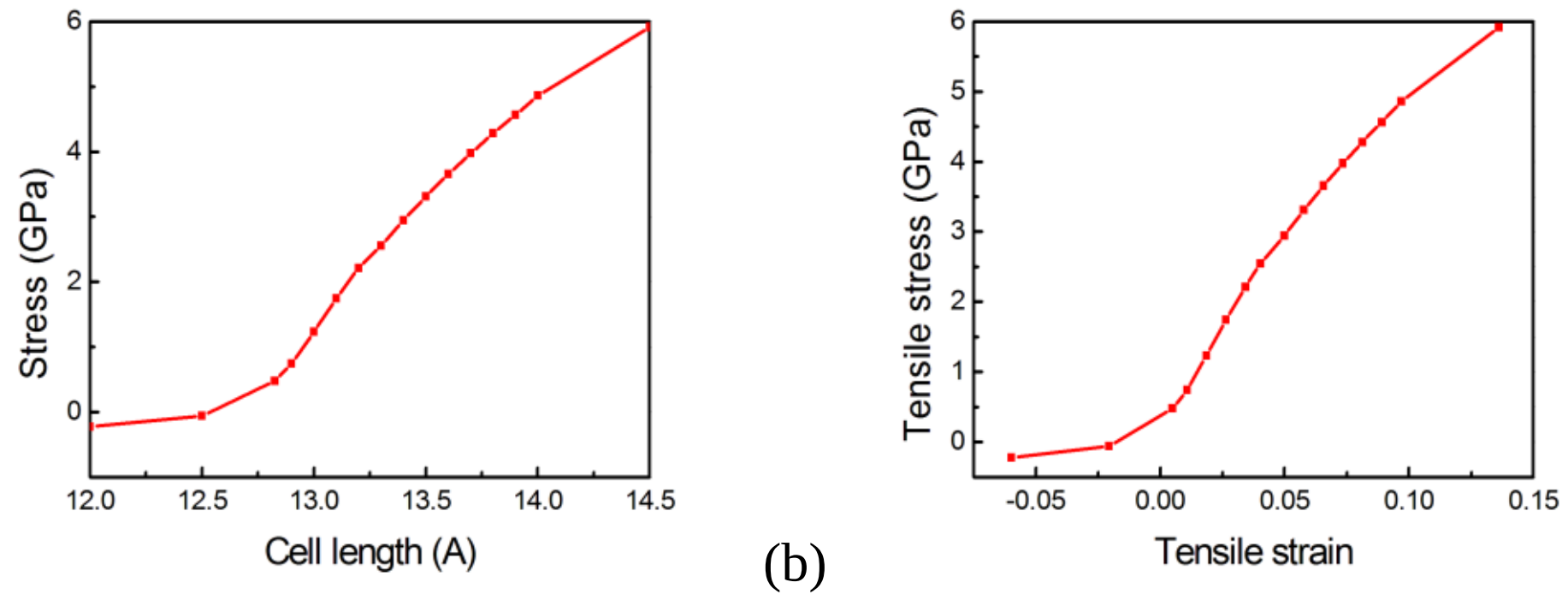

(a) (b)

Fig.4 The tensile stress of cumulene as a function of the cell length (a) or strain (b).

## 4.2 The mechanics properties of polyyne

In this subsection, we move to the discussion of the mechanical properties of polyyne. We first calculate the tensile and bond stiffness of polyyne using the same method as in cumulene. The results for the dependence of the total energy and band gap on the strain are shown in Fig.5. The tensile stiffness can be obtained from Fig.5(a) which turns out to be 94.939eV/A, well coinciding with other results, e.g., 95.56eV/A in Ref.[12]. From Fig.5(b), it can be seen that the band gap decreases with the increase of the bond strain, or the bond length. We obtain chemical bond stiffness $K$ in polyyne to be 101.42GPa based on the method for cumulene.

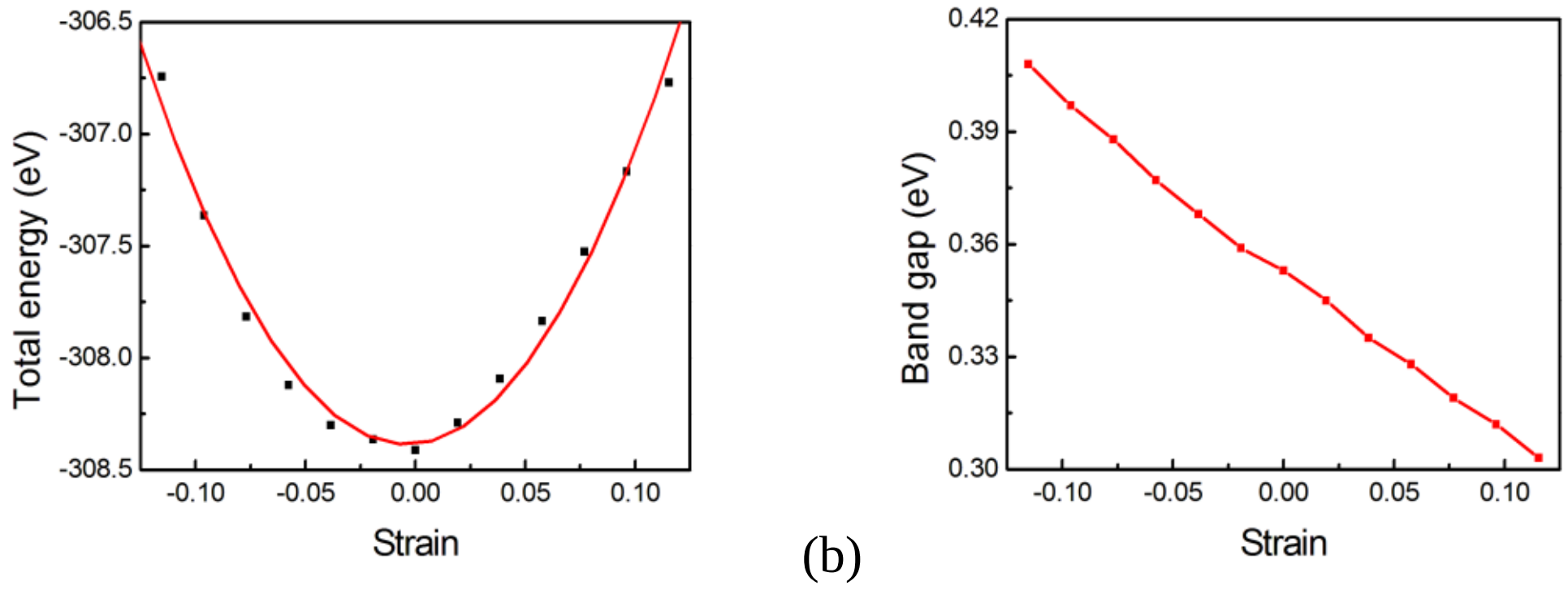

(a) (b)

Fig.5 The total energy (a), band gap (b) as a function of strain.

For polyyne, we are specially interested in the change of the mechanical properties with the fractional coordinate of the mid-carbon atom. Namely, the fractional coordinate measures the relative position of the mid-carbon atom with respect to the two fixed ends in the unit cell of polyyne (here the length is 0.2566nm). In doing this, we perform the calculation without geometry optimization (otherwise with the application of geometry optimization, polyyne will automatically change into cumulene). The results of the total energy and the band gap as a function of fractional coordinate are shown in Fig.6. We can see from Fig.6 that both the total energy and the band gap decrease monotonically with the increase of the fractional coordinate of the mid-carbon atom from 0.40 to 0.50. Moreover, the linear dependence of the band gap on the fractional coordinate can be observed. When the fractional coordinate of the mid-carbon atom reaches 0.5, i.e., polyyne turns into cumulene, the total energy reaches its minimum value and at the same time, the band gap is zero by extrapolation. We may conclude that cumulene is a conductor while polyyne is a semiconductor. Besides, from the point of view of the total energy, cumulene seems to be more stable than polyyne while keeping the cell size unchangeable. If the cell size is optimized, polyyne would be more stable than cumulene, as narrated in “crystal binding energy and formation energy of carbyne ”.

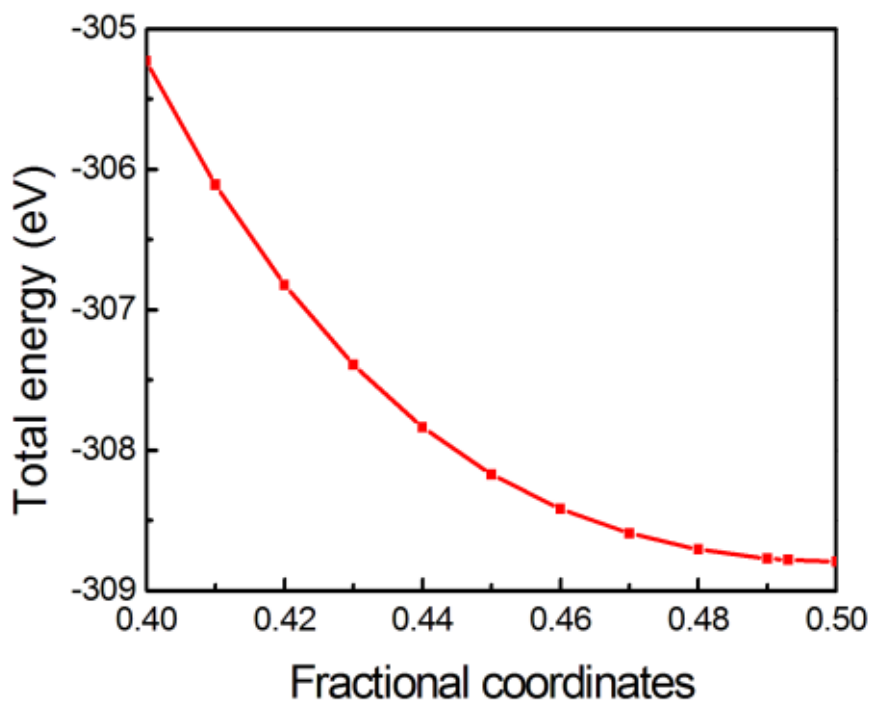

(a)

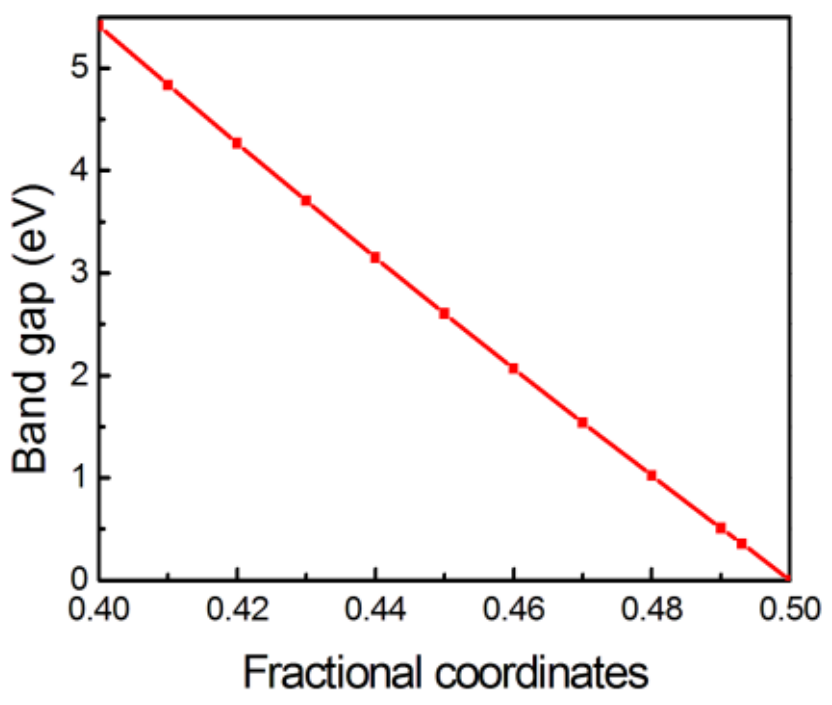

(b)

Fig.6 The total energy (a), and the band gap (b) of polyyne as a function of the fractional coordinate without geometry optimization.

A similar calculation starting from the side of cumulene without geometry optimization can also be performed. The result shows little quantitative difference compared with the case of polynne (see Fig.7). We can see from Fig.7 that both the total energy and the band gap decrease monotonically with the increase of the fractional coordinate of the mid-carbon atom from 0.40 to 0.50. Moreover, the linear dependence of the band gap on the fractional coordinate can be observed. When the fractional coordinate of the mid-carbon atom is 0.4, the length between the mid-carbon atom and the end-carbon atom is 0.15396nm, very similar to the shortest-bond length in diamond (0.1544nm). It is well known that diamond is semiconductor, and its band gap is 5.47 eV. When the fractional coordinate of the mid-carbon atom is 0.4, the band gap is about 5.415eV in the polyyne, as shown in Fig.6 (b).

(a) 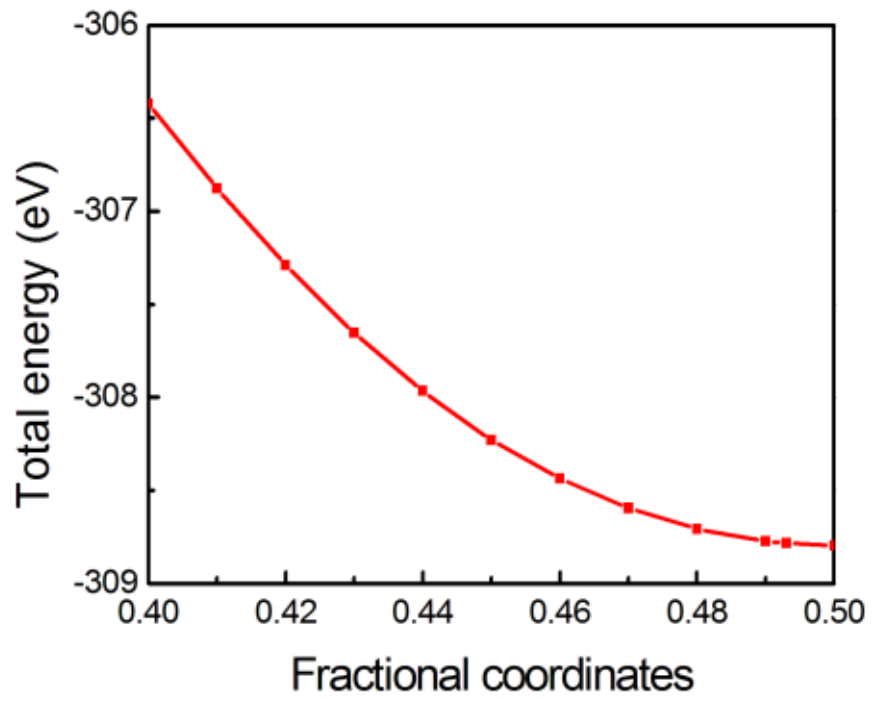

(b) 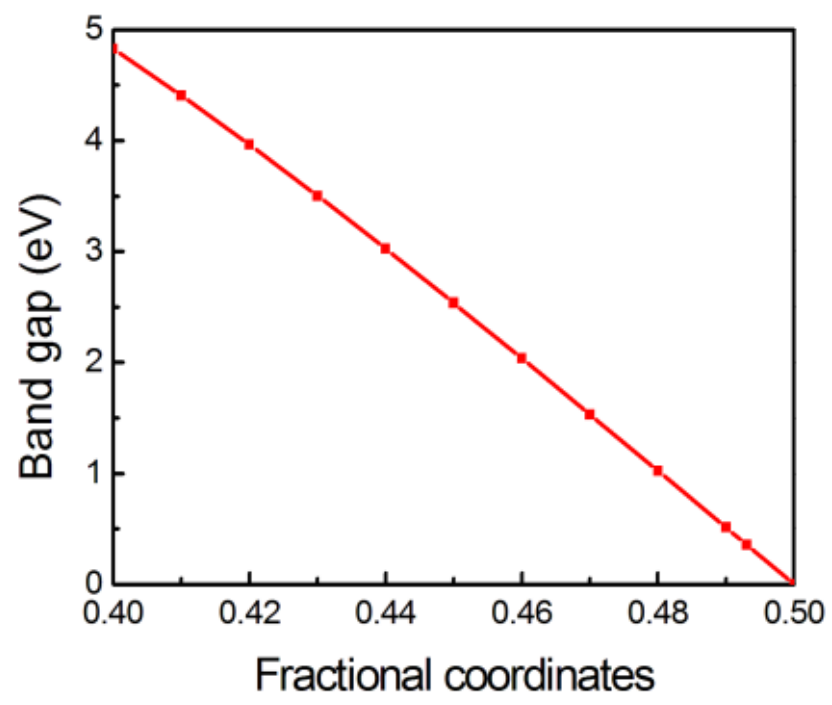

Fig.7 The total energy (a), and band gap (b) of cumulene as a function of the fractional coordinate without geometry optimization.

## 5. Effect of the supercell number

In this section, we discuss the impact of the adopting supercell number on the properties of polyyne and cumulene in our first-principles calculation. The results on polyyne are shown first, where the supercell number varies from 1×1×1 to 1×1×9. It is no surprise that the cell length linearly increases with the supercell number N (Fig.8(a)). However, there are plain even-odd oscillations in the band gap, bulk modulus and the final energy with respect to the change of the supercell number N (in Figs.8(b), (c), (d)). In more detail, at N=5, the final energy reaches its global minimum, while at the same time, the band gap and the bulk modulus reach their maximums. At N=4, the situation is reversed. We can conclude that the system is the most stable when the supercell number N=5.

The results for the effect of the supercell number in cumulene after geometry optimization are summarized in Fig.9. The size of the supercell also changes from 1×1×1 to 1×1×9. Similar even-odd oscillations can be found. Based on the same arguments, we see that for cumulene, N=5 is also the most stable.

(a) 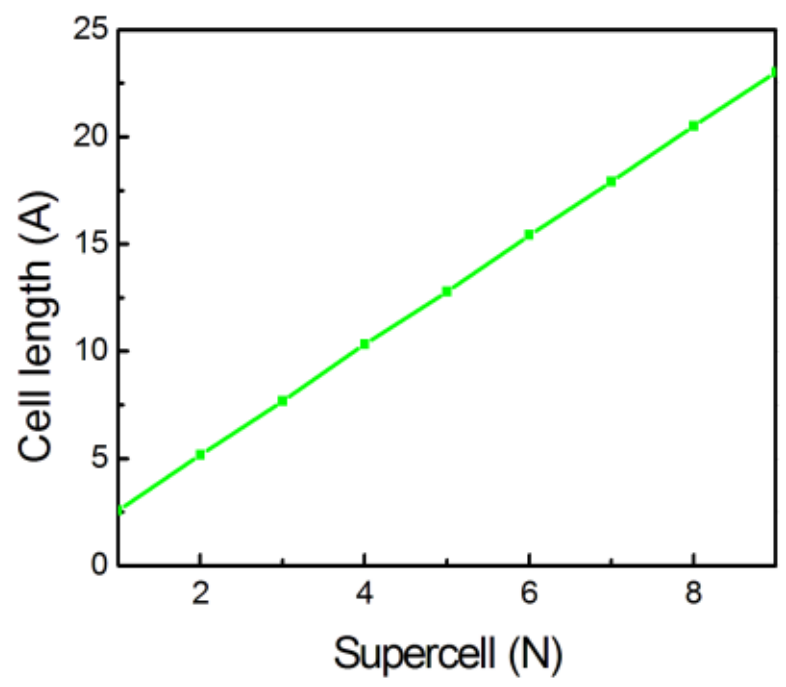

(b) 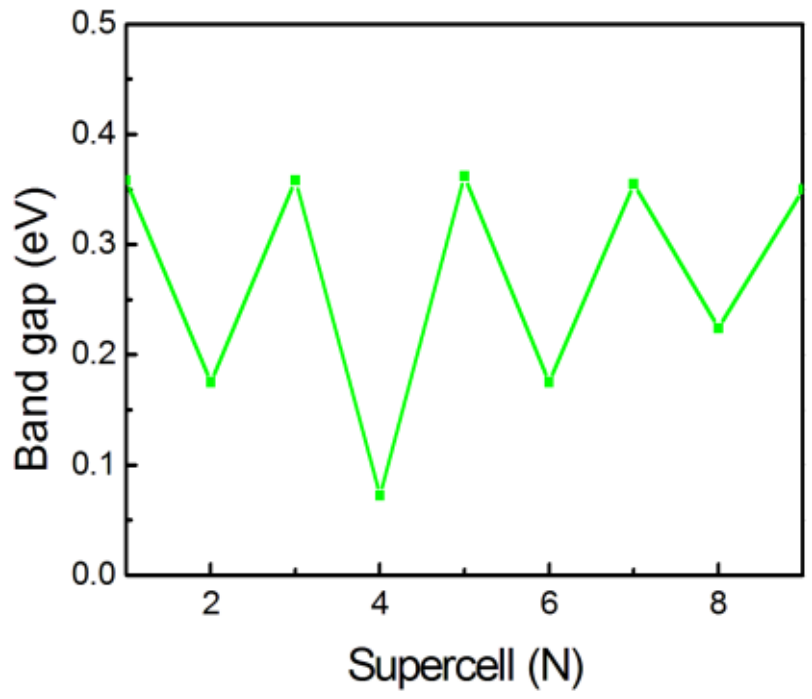

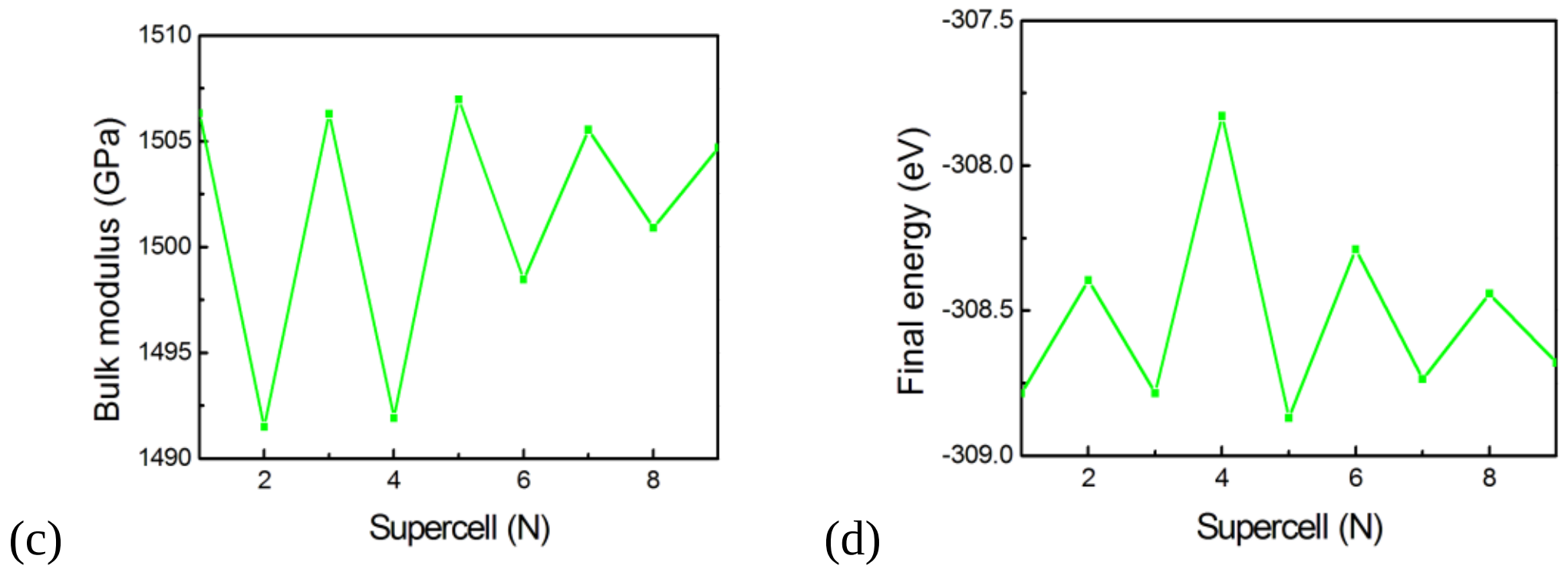

Fig.8 The characteristics such as the cell length (a), band gap (b), bulk modulus (c), and final energy (d) in polyyne changing with the supercell number.

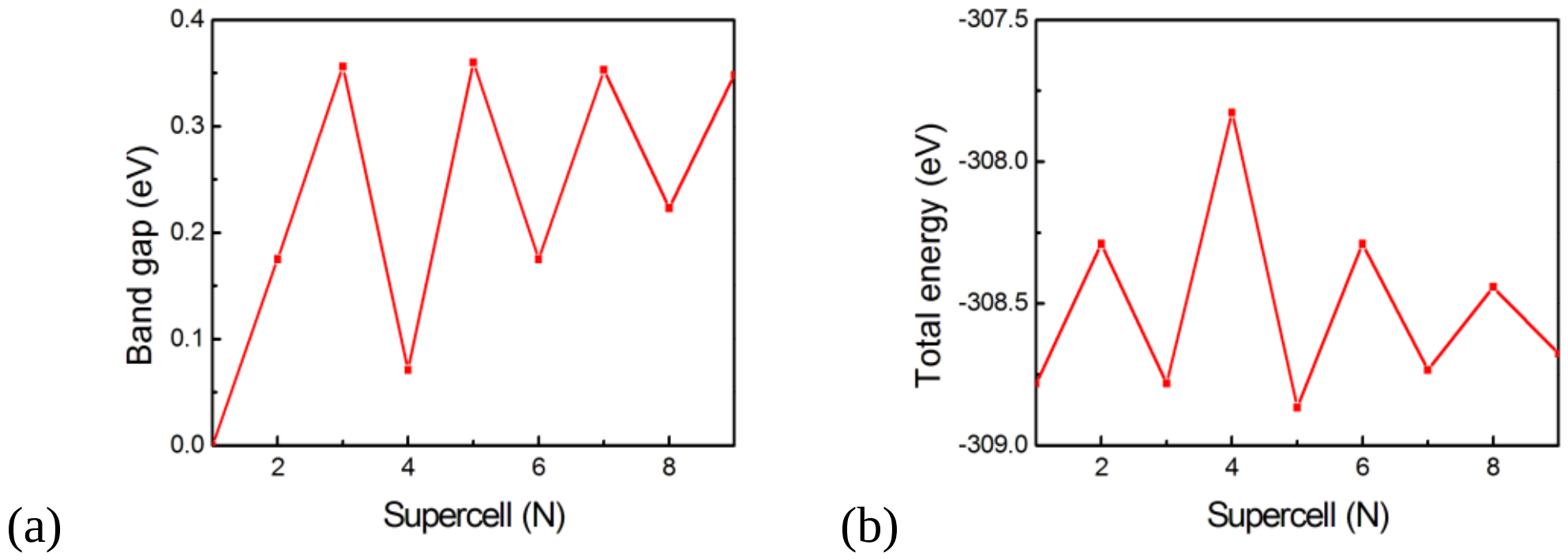

Fig.9 The changes of the band gap (a), and total energy (b) in cumulene with the supercell number.

In the following calculation, we set the supercell of polyyne to be 1×1×5 because it is the most stable configuration. And then we study the deformation under the unidirectional tensile stress. The calculation parameters are set as: c=2.566×5=12.830 Å, GGA-PW91, ultrafine, modulus-conserving, 72 empty bands while we keep the fractional coordinates of the mid-carbon atoms, three right angles and a/b cell length unchanged. When the pull value increases from 0 to 2.5GPa, we obtain the band gap, bulk modulus, final energy and strain of polyyne after geometry optimization.

The numerical results about the band gap, bulk modulus, final energy and strain of polyyne changing with tensile stress are shown in Fig.10. From Fig.10, we can see that with the rise of tensile stress, the band gap (Fig.10(a)), bulk modulus (Fig.10(b)) and final energy (Fig.10(c)) decrease linearly, while the strain increases (Fig.10(d)). From Fig.10(d), we can obtain the Young's modulus as 60.06GPa for the polyyne supercell, equal value to the unit cell. From the point of view of energy, the bigger the tensile pressure is, the more stable the polyyne supercell is (shown in Fig.10(c)).

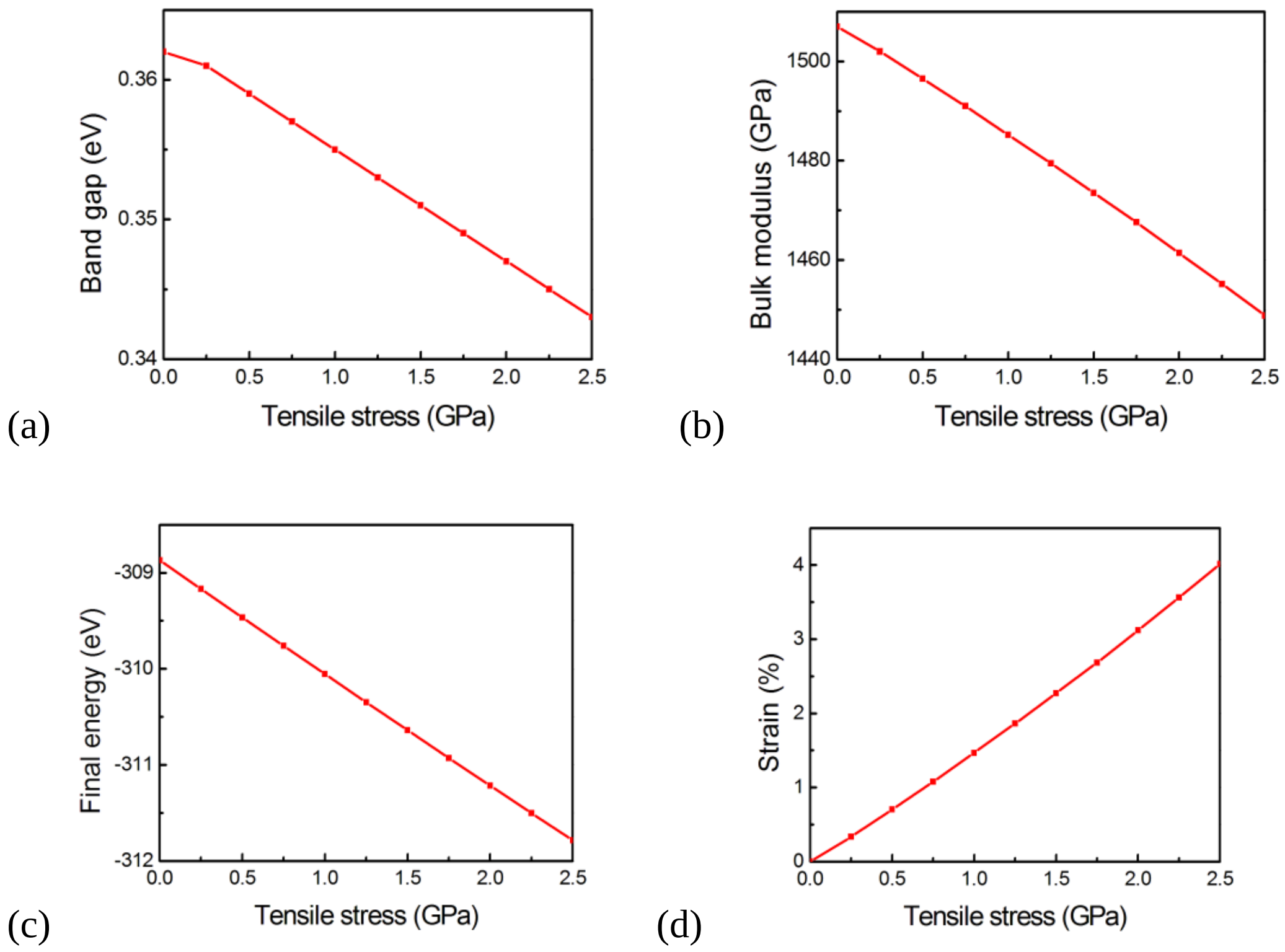

Fig.10 The characteristics such as the band gap (a), bulk modulus (b), final energy (c) and strain (d) of polyyne changing with tensile stress.

Next, similar to the discussion for cumulene in Sec.4.2, we study the response of polynne under the application of strain. The calculation parameters are set as: $c_0$=2.566×5=12.830 Å (c=12.770581 Å after geometry optimization), GGA-PW91, ultrafine, modulus-conserving, 72 empty bands and at the same time while we keep the fractional coordinates of the mid-carbon atoms, three right angles and a/b cell length unchanged. With $c_0$ length changed from 12 to 14 Å under the strain, we obtain the corresponding values of the cell length, band gap, stress and final energy after geometry optimization, as shown in Fig.11. From Fig.11(a) and (b), the linear dependence of the cell length and the band gap on the strain (up to an increment of 15%) can be observed. On the other hand, the deviation from the linear dependence between stress and the applied strain is clear in Fig.11(c). Furthermore, in Fig.11(d) for the final energy, we can see that it is a non-monotonic function of strain and has the minimum when no strain is applied, which corresponds to c=12.770581 Å after geometry optimization.

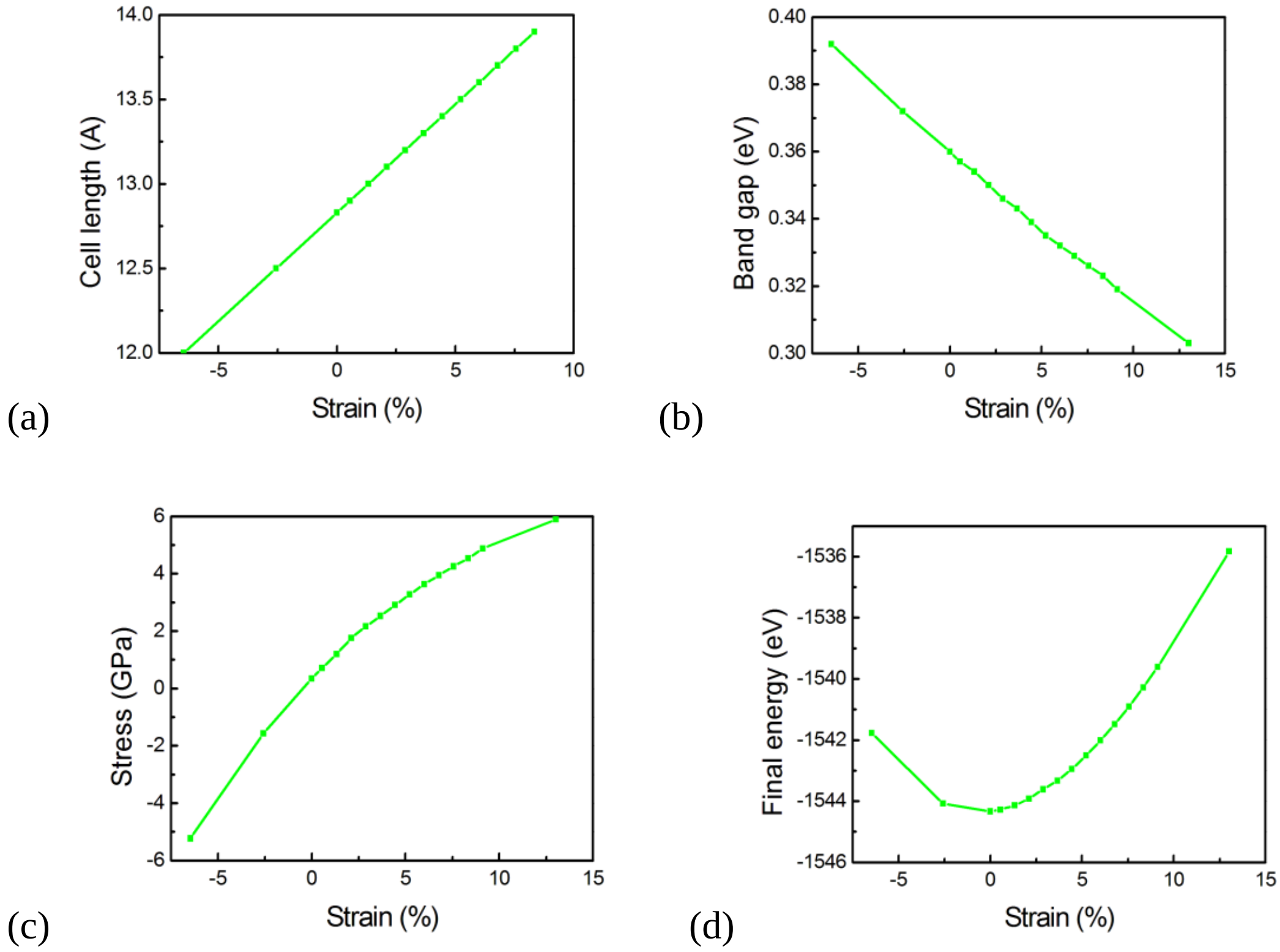

Fig.11 The characteristics such as the cell length (a), band gap (b), final energy (c) and stress (d ) of polyyne changing with strain.

# 6. Electronic structure, optical and phonon properties of cumulene

## 6.1 Electronic structure of cumulene

The band structure along high-symmetry points in the Brillouin zone and the density of states (DOS) are shown in Fig.12. No band gap is observed which indicates cumulene is a conductor. The total DOS is presented in Fig.12(b). There are three parts in the valence-band regime, i.e., the lowest region about -16~-12.5eV, the lower region about -12.5eV~-7.5eV, and the upper region about -5eV~0eV, which mainly come from C 2*s*, C 2*s*+2*p*, C 2*p* electrons, respectively.

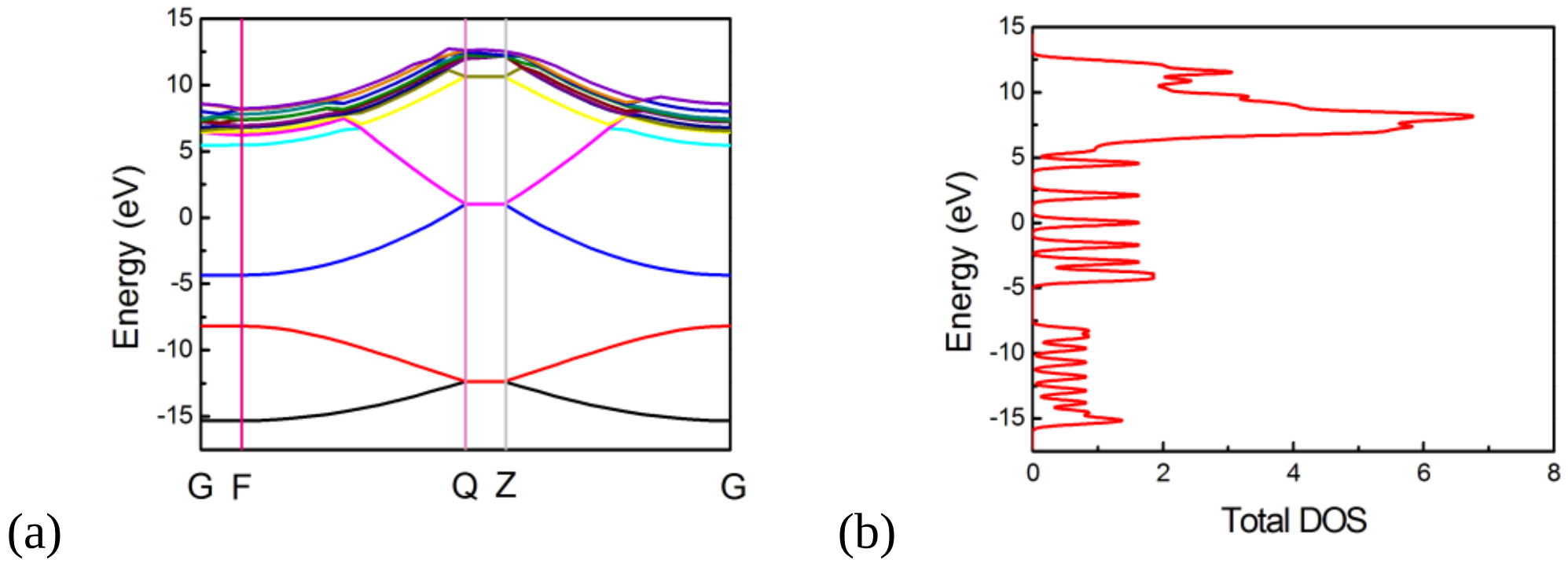

Fig.12 The band structure along high-symmetry points in the Brillouin zone (a), and the total DOS (b) of cumulene.

## 6.2 Optical properties

The complex dielectric function ε contains a real part $\varepsilon_1$ and an imaginary part $\varepsilon_2$. The imaginary part $\varepsilon_2$, which is related to the real part of optical conductivity, can be calculated from the band structure directly by taking into account inter-band transitions, while the real part $\varepsilon_1$ can then be obtained by employing the Kramers-Kronig relations [14]. Figure 13 shows the average and the orientation dependence of the dielectric function as a function of photon energy obtained by the GGA. In Fig.13, (001) is the direction along the chain, and (010), (100) are transverse directions. From Fig.13, we can determine the static dielectric constant ($\varepsilon_1$ at E=0) of cumulene as 1.02, and the maximums of the real part and the imaginary part of the dielectric function, which are 1.155 and 0.213, appear at 9.47eV and 9.79eV, respectively. Moreover, the anisotropy of the dielectric function among (001), (010), and (100) is clearly observed, while that along (010) or (100) is almost same to the mean value, consistent with the one dimensional nature of the carbon chains.

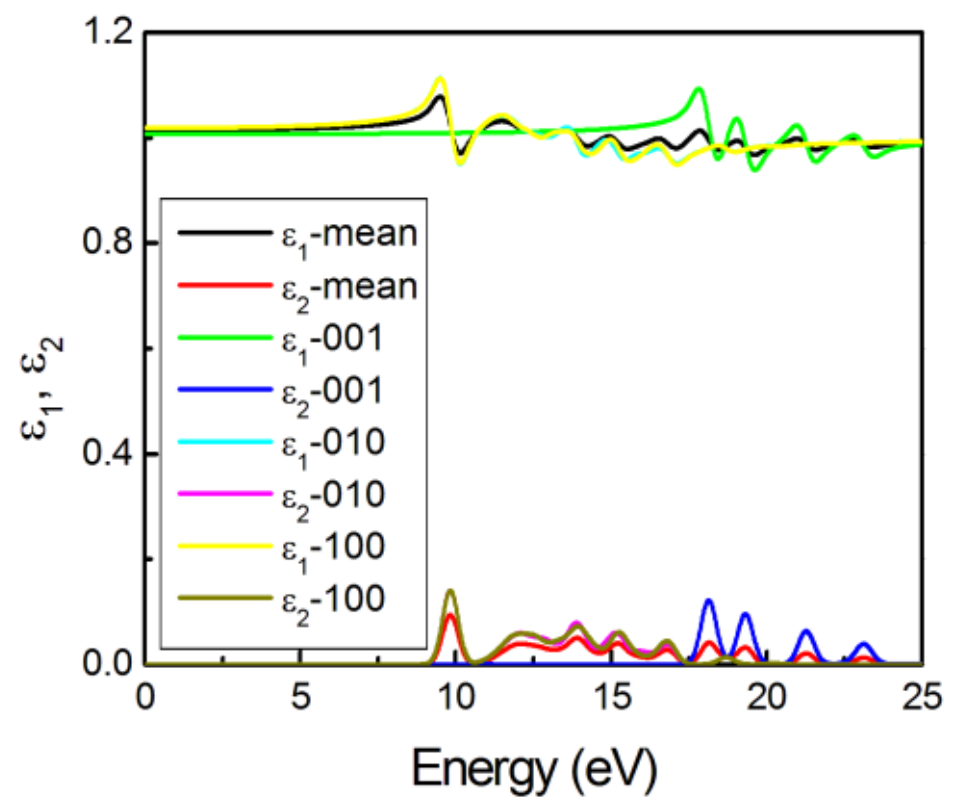

Fig.13 The average and the orientation dependence of the complex dielectric function for cumulene.

## 6.3 Phonon properties

The low-energy part of the phonon dispersion of cumulene and its DOS obtained by GGA-PW91 are shown in Fig.14. The Bradley-Cracknell notation is used for the high-symmetry points, e.g., G=(0, 0, 0), Q=(0.0, 0.0, 0.47), and Z=(0.0, 0.0, 0.5). Since each unit cell of cumulene contains two carbon atoms, there are six vibration modes including three acoustic and three optical ones. As shown in Fig.14, the modes are mixed and there is no phonon band gap. If the phonon dispersion of one material has no negative frequency, it illustrates that it is kinetically stable [15]. For cumulene, there are some parts with negative frequencies with the lowest value down to -3.817THz, which indicates that cumulene is kinetically unstable and it could be unstable at room temperature and normal pressure.

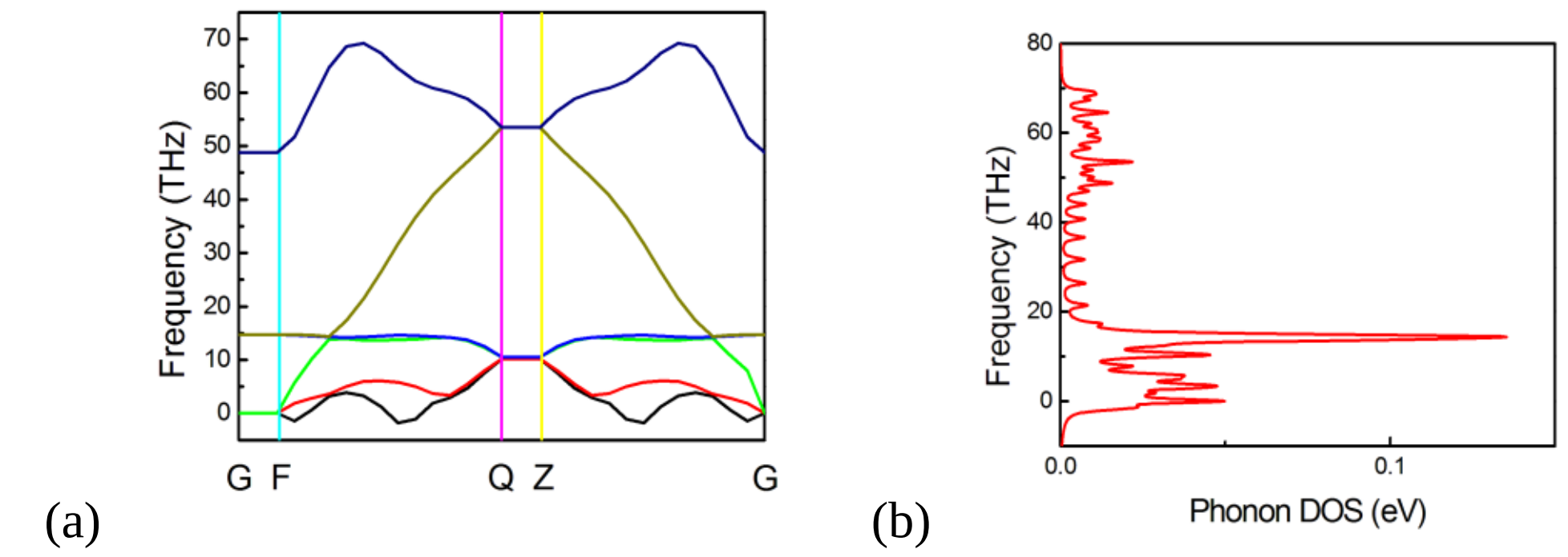

Fig.14 The low-energy part of the phonon dispersion (a) and the phonon DOS (b) of cumulene obtained by GGA-PW91.

Some thermodynamic quantities obtained by GGA-PW91, including the enthalpy (which is just the summation of the Gibbs free energy and a term given by the product of temperature and entropy, T*entropy), Gibbs free energy, entropy, and the heat capacity of the cumulene, are shown in Fig.15. From Fig.15(a), it can be seen that the internal energy of the cumulene raises slowly with temperature from 20 to 200K, and then increases nearly linearly over 200K. Entropy increases with temperature. Since the ions polarize each other when they close each other, the relationship line between entropy and temperature presents certain bending. The free energy decreases with temperature, and the relationship curve between free energy and temperature presents concave downward. With rise of temperature, the heat capacity (shown in Fig.15(b)) first increases fast, and then slows down, gradually approaching the Dulong-Petit limit, i.e., $C$=3$NkV$, in the high-temperature regime.

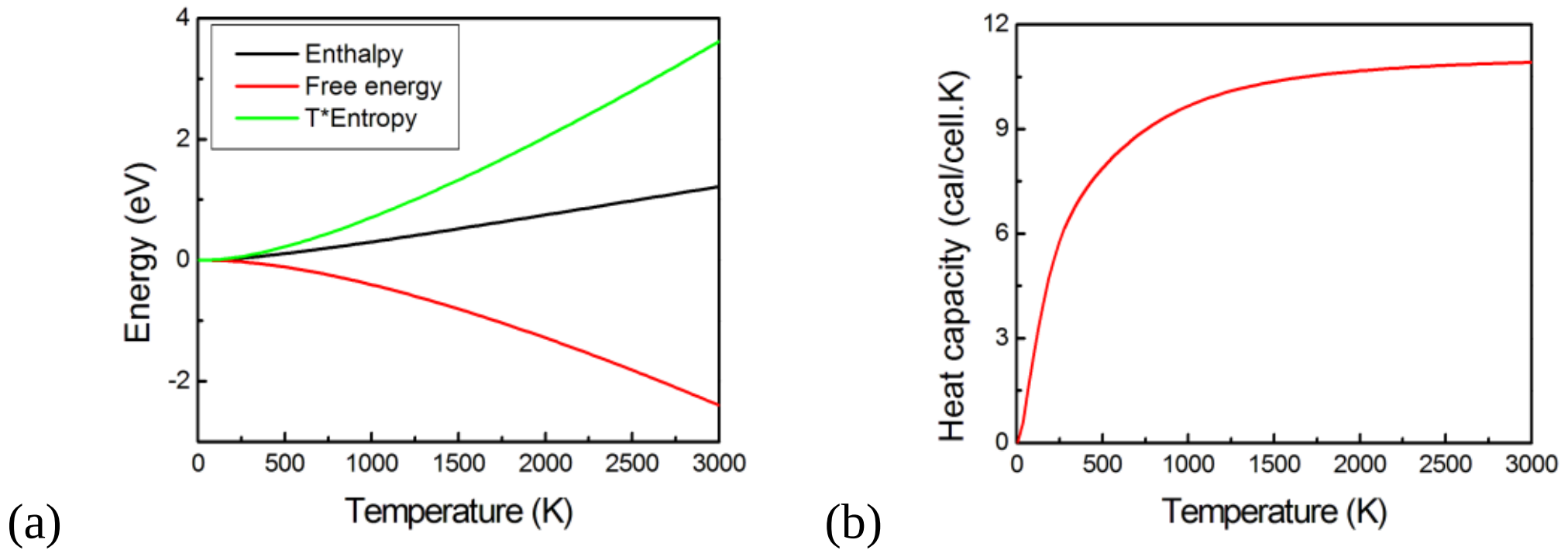

Fig.15 Some thermodynamic quantities (a) and the heat capacity (b) of the cumulene obtained by GGA-PW (the zero point energy is 0.2599eV ).

# 7. Electronic structure and optical properties of Polyyne

## 7.1 Electronic structure

The band structure along high-symmetry points in the Brillouin zone and the DOS are shown in Fig. 16. The tiny band gap with the value of 0.37 eV is observed (Fig.16(a)), which indicates polyyne is a semi-conductor, coincided with the experimental results [16], higher than the other theoretical one, 0eV in Ref. [7]. The total DOS is presented in Fig.16(b). There are three parts in the valence bands, *i.e.*, the lowest region about -20~-15eV, the lower region about -15eV~-10eV, and the upper region about -6eV~0eV, which mainly come from C 2*s*, C 2*s*+2*p*, C 2*p* electrons, respectively.

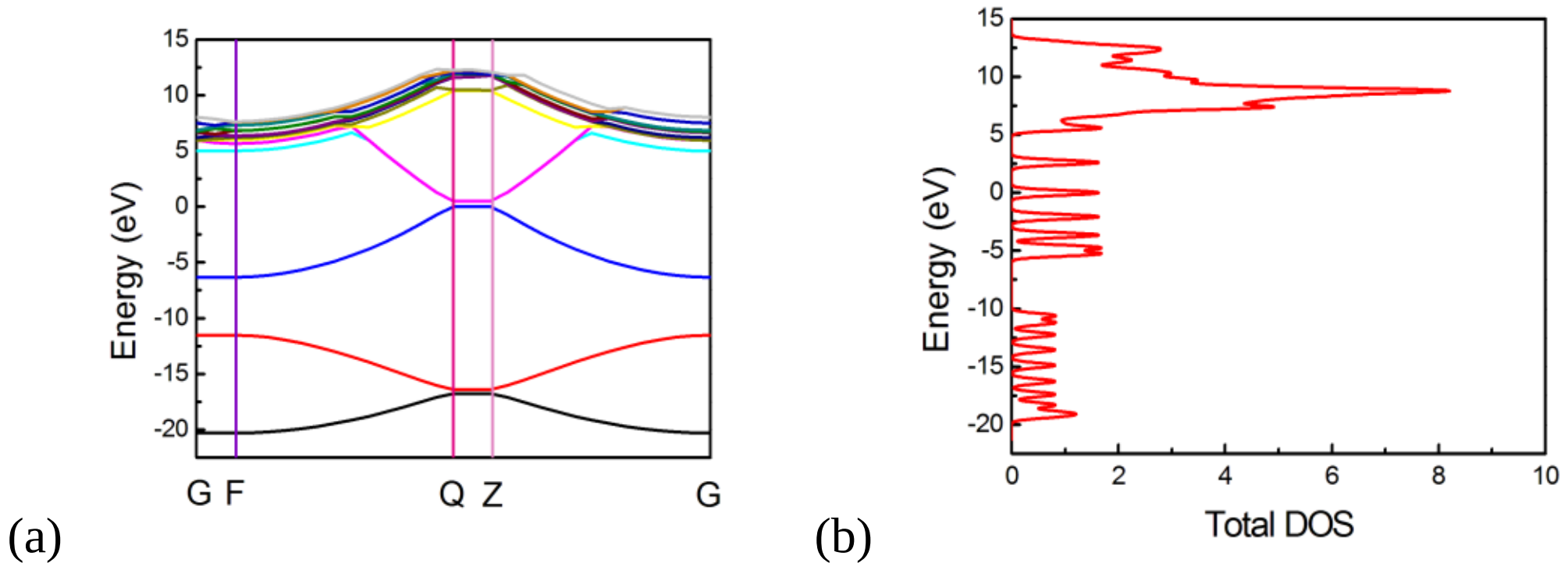

Fig.16 The band structure along high-symmetry points in the Brillouin zone (a) and the total DOS (b) of polyyne.

### 7.2 Optical properties

Figures 17 show the complex dielectric function as a function of photon energy obtained by GGA. We use three supercells, i.e., 1×1×1, 1×1×5 and 1×1×9, for polyyene. The dielectric function for different supercell numbers are shown in Fig.17. We can see that with the increase of N, the dielectric function shifts toward lower energy as a whole, yet with no significant change. For 1×1×1 supercell, we can see that the static dielectric constant ($\varepsilon_1$ at E=0) is 1.03, and the maximums of the real part and the imaginary part of the dielectric function are 1.15 and 0.2, respectively. Their corresponding energy positions, around 2.5eV, are almost identical.

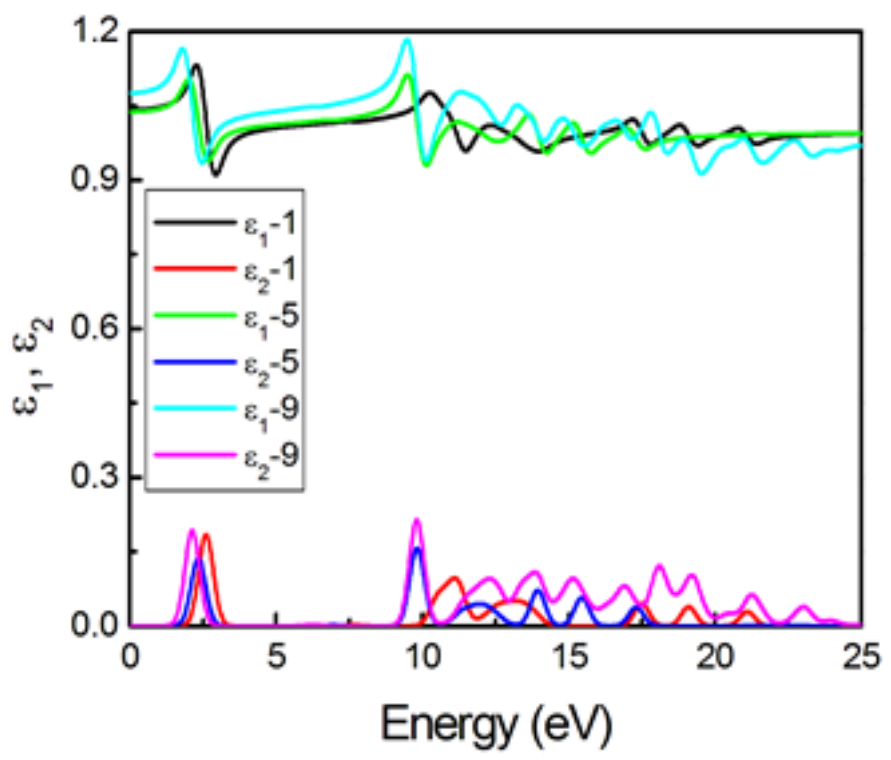

Fig.17 The complex dielectric functions in different polyyne supercells.

## 8. Conclusion

We calculate the mechanical, electronic band structure, optical and phonon properties of the two kinds of carbyne, i.e., cumulene and polyyne, by the first principles. The mechanical properties including the tensile stiffness, bond stiffness and Young's modulus show similar behaviors; while quantitatively, polyyne turns out to be more stable and harder than cumulene. The results on the electronic band structure show that cumulene is a conductor while polyyne is a semiconductor. The supercell calculation suggests that carbyne is the most stable when the supercell number N=5. The dielectric function of carbynes varied in different direction, which is consistent with the one dimensional nature of the carbon chains. The presence of the imaginary frequencies in the phonon dispersion of cumulene indicates that at room temperature and normal press, cumulene might be unstable.

## Acknowledgments

This work was supported by the National Natural Science Foundation of China (Grant no. 11672087, 11474136), the Strategic Programs for Innovative Research (SPIRE), the Computational Materials Science Initiative (CMSI) and the Yukawa International Program for Quark-Hadron Sciences at YITP, Kyoto University.

## References:

[1] Belenkov E A, Mavrinsky V V 2008 *Crystallography Reports* **53**(1) 83-87

[2] Khoo K H, Neaton J B, Son Y W, Cohen M L, Louie S G 2008 *Nano Letters* **8**(9) 2900-2905

[3] Zanolli Z, Onida G, Charlier J C 2010 *ACS Nano* **4**(9) 5174-5180

[4] Zeng M G, Shen L, Cai Y Q, Sha Z D, Feng Y P 2010 *Applied Physics Letters*

**96**(4) 042104
[5] Akdim B, Pachter R 2011 *ACS Nano* **5**(3) 1769-1774
[6] Artyukhov V I, Liu M, Yakobson B I 2014 *Nano Letters* **14**(8) 4224-4229
[7] Zhang Y Z, Su Y J, Wang L, Kong E S, Chen X S, Zhang Y F 2011 *Nanoscale Research Letters* **6**(1) 577
[8] Sorokin P B, Lee H, Antipina L Y, Singh A K, Yakobson B I 2011 *Nano Letters* **11**(7) 2660-2665
[9] Cahangirov S, Topsakal M, Ciraci S 2010 *Physical Review B* **82**(19) 195444
[10] Webster A 1980 *Monthly Notices of the Royal Astronomical Society* **192**(1) 7P-9P
[11] Chalifoux W A, Tykwinski R R 2010 *Nature Chemistry* **2**(11) 967-971
[12] Liu M J, Artyukhov V I, Lee H, Xu F B, Yakobson B I 2013 *Acs Nano* **7**(11) 10075-10082
[13] Segall M D, Lindan P J D, Probert M J, Pickard C J, Hasnip P J, Clark S J, Payne M C 2002 *Journal of Physics: Condensed Matter* **14**(11) 2717
[14] Wang J, Li H P, Stevens R 1992 *Journal of Materials Science* **27**(20) 5397-5430
[15] Zhang Q, Zhang H, Cheng X L 2018 *Chinese Physics B* **27**(2) 027301
[16] Kudryavtsev Y P, Evsyukov S E, Guseva M B, Babaev V G, Khvostov V V 1993 *Russian* Chemical *Bulletin* **42**(3) 399-413